# MESOSCOPIC STUDY ON HISTORIC MASONRY

**J. Šejnoha[*,1,2], M. Šejnoha[1,2], J. Zeman[1], J. Sýkora[1] and J. Vorel[1]**

[1] Department of Mechanics, Faculty of Civil Engineering, CTU in Prague,

Thákurova 7, 166 29 Praha 6, Czech Republic

[2] Centre for Integrated Design of Advanced Structures, CTU in Prague,

Thákurova 7, 166 29 Praha 6, Czech Republic

## ABSTRACT

This paper presents a comprehensive approach to the evaluation of macroscopic material parameters for natural stone and quarry masonry. To that end, a reliable non-linear material model on a meso-scale is developed to cover the random arrangement of stone blocks and quasi-brittle behaviour of both basic components, as well as the impaired cohesion and tensile strength on the interface between the blocks and mortar joints. The paper thus interrelates the following three problems: (i) definition of a suitable periodic unit cell (PUC) representing a particular masonry structure; (ii) derivation of material parameters of individual constituents either experimentally or running a mixed numerical-experimental problem; (iii) assessment of the macroscopic material parameters including the tensile and compressive strengths and fracture energy.

[*] Correspondence to: J. Šejnoha, Department of Mechanics, Faculty of Civil Engineering, CTU in Prague, Thákurova 7, 166 29 Praha 6, Czech Republic. E-mail: sejnoha@fsv.cvut.cz

# 1 INTRODUCTION

Masonry structures have been extensively used in the whole history of mankind, mainly due to a wide availability of the material as well as its good mechanical properties. In the past, the design and construction of these structures were based on a balanced combination of experience and trial-and-error methods. Even nowadays, in spite of the progress in constitutive and numerical modelling, the fact remains that the engineering approach to these structures builds upon a number of simplifying assumptions and phenomenological relations, see e.g. (Lourenço, 2002) for further discussion.

However, the limitations and even the inadequacy of such assumptions call for more advanced constitutive models to provide a solid prediction of the mechanical response of masonry structures. Material models, already developed in the distant past, can be broadly classified into two main categories. The first category is characterized by closed-form macroscopic constitutive laws (Pande et al., 1989; Lourenço et al.; 1997, Papa and Nappi; 1997). In the second category, a mesoscopic approach is used by interpreting each of the constituents (i.e. stone blocks and mortar joints) as an individual body, endowed with specific geometric and material properties (Hart et al., 1988; Lourenço and Rots, 1997; Giambanco et al., 2001). This category also includes a novel methodology of modelling masonry on a mesostructural scale using the partition of unity concept of finite element shape functions presented by De Proft and Sluys (2005). While big savings on a computational effort constitute the main benefit of the first concept, the lack of a well-founded mechanical basis, however, could be one of its crucial drawbacks.

Most modern approaches are therefore based on the conjunction of both concepts. They may be regarded as essentially continuous models enriched with a deep micromechanical insight, and could be referred to as multi-scale approaches, see (Anthoine, 1995, 1997; Phillips, 1998; Smit et al.,1998; Michel et al., 1999, Kouznetsova et al., 2001; Massart et al., 2007) and references therein. Therein, the so-called standard multi-scale method that intrinsically couples both the mesoscopic and macroscopic scales of representation is introduced. However, its application is restricted to situations in which the principle of the separation of scales is preserved. It means that the mesoscopic characteristic length is much smaller than the length scale associated with the variations of fields on the macro-scale. Masonry structures are, unfortunately, a typical example where such an assumption may fail. A promising solution to this problem is introduced in (Massart et al., 2007) where a multi-scale approach relying on the first-order homogenization framework is enhanced in such a way that both scales are fully coupled in the entire structural computation and a finite width of the damage band model is added to the macroscopic description in order to allow the treatment of macroscopic localization resulting from a damage growth in the constituents.

Regardless of the material systems being investigated, the above-mentioned solution strategy was applied exclusively to two-dimensional structures. When referring specifically to masonry structures, only simple 2D panels made up of regular brickwork were analyzed. A straightforward extension of fully coupled multi-scale strategies to real, generally large three-dimensional, masonry structures still appears computationally unfeasible. It therefore follows that, in typical applications of engineering practice, the analysis on individual scales would likely be kept entirely independent. An assessment of three masonry towers subjected to different loading conditions, studied in (Carpinteri et al., 2006), supports this opinion. Such a methodology, promoted by a recent need for a repair and rehabilitation of Charles Bridge in Prague, is also adopted in the present contribution. While a fully three-dimensional structural (macroscopic) nonlinear analysis is performed to provide estimates of the current state of damage and to unveil the main sources of possible failure of complex historical structures, see e.g. Fig. 1(b), a detailed independent mesoscopic analysis is used for the predictions of homogenized (macroscopic) material parameters. As a result of such a sequential multi-scale computational approach the timetable, computational cost and reliability of the expected results can be well balanced.

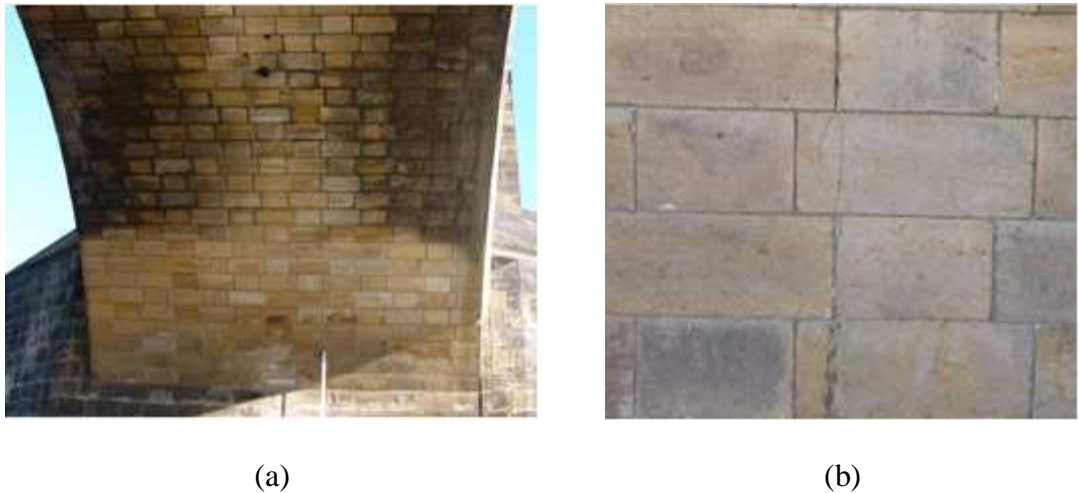

(a)          (b)

Figure 1: a) A view of a typical bridge arch showing a regular arrangement of stones in masonry, b) a crack running both across stone and along a head joint.

Once the homogenized material parameters are available, the subsequent macroscopic structural analysis becomes a relatively standard task. We therefore limit our attention to the first step concerned with the formulation of a reliable method estimating material parameters characterizing the effective mechanical properties, including fracture energy, of the masonry to be analyzed. Special attention is focused on the homogenization of quarry masonry and masonry with an irregular geometry.

In particular, the solution procedure relies on a well founded first-order homogenization strategy outlined, e.g. by (Michel et al., 1999). In view of historical masonry structures, its application requires the solution of three specific problems:

- Formulation of a periodic representative volume element (RVE) for a masonry structure with a disordered geometry. The respective PUC results from matching geometrical statistics related to the original structure and the idealized cell, respectively. A brief review of this concept is outlined in Section 2.
- Derivation of local material parameters that appear in the selected material model. Here, the constitutive model implemented in the ATENA commercial computer code (Červenka et al., 2002), which allows us to treat both basic components as quasi-brittle materials, is adopted. While material parameters of individual constituents are easily provided by standard experiments, see e.g. (Novák et al., 2006), the behaviour along their interfaces, crucial for reliable estimates of the homogenized response, is likely to be predicted from a combined numerical – experimental analysis. This new insight into the mesoscopic modelling of historic masonry is presented in Section 3.
- Nonlinear homogenization at the level of the PUC providing the desired homogenized effective material properties such as elastic stiffness, the macroscopic tensile and compressive strengths and in particular the macroscopic fracture energy. This step is addressed in Section 4. A homogenization-based approach to the prediction of the macroscopic fracture energy is further validated through an independent study that draws on a series of numerical representations of the macroscopic wedge splitting test assuming specimens of variable ligament lengths. Further applications of both linear and nonlinear analysis at the level of the PUC can be found in, e.g. (Anthoine, 1995, 1997; Massart, 2007).

In this work, symbols *a*, ***a*** and **A** denote a scalar, a column vector and a matrix, respectively. Moreover, the standard Voight notation is employed for the representation of symmetric second- and fourth-order tensors, see e.g. (Bittnar and Šejnoha, 1996).

## 2 Mesoscopic geometrical modelling: definition of PUC

When adopting the sequential multi-scale modelling approach, the specification of the geometry on the mesoscale is provided by a notion of the Representative Volume Element (RVE), which corresponds to the statistically equivalent sample of the analyzed part of the structure. In the context of historical masonry structures, three typical material morphologies can be identified:

- *Regular periodic* stone masonry, Fig. 2(a), characterized by a *Periodic Unit Cell* with geometrical parameters specified by on-site measured parameters
- Masonry with a *non-periodic* arrangement of individual blocks, Fig. 2(b). In this case, the geometry of the real-world material is specified by a *Statistically Equivalent Periodic Unit Cell*, derived by a methodology proposed by Povirk in (Povirk, 1995) and further extended by Zeman and Sejnoha (2001,2007). The procedure is based upon replacing the complex and possibly non-periodic structure by a simpler PUC, which still optimally resembles the original material in the sense of selected geometrical statistical descriptors. Note that this approach is well-suited for the current application as it allows a direct use of digitized photographs of real-world structures; see (Cluni and Gusella, 2004; Zeman and Šejnoha, 2007) for more details related to the application to irregular masonry structures.

- *Irregular filling (quarry masonry)* of selected parts of the bridge. In this case, a representative volume element is an expert estimate based on the elementary statistical characterization obtained from dug holes, such as size distribution of individual stones and basic shape characterization. An example of such a structure is shown in Fig. 2(c).

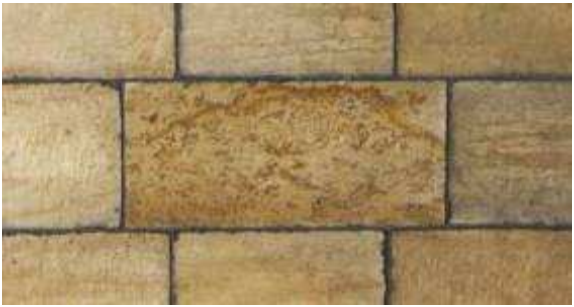 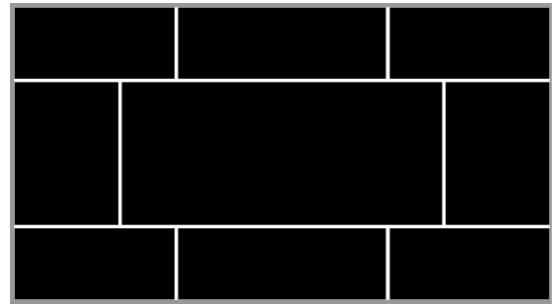

(a)

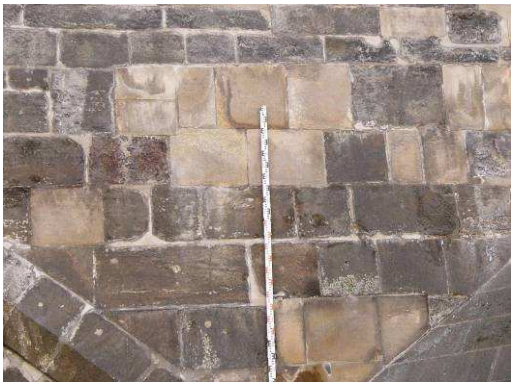 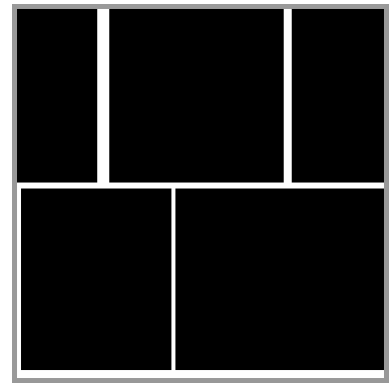

(b)

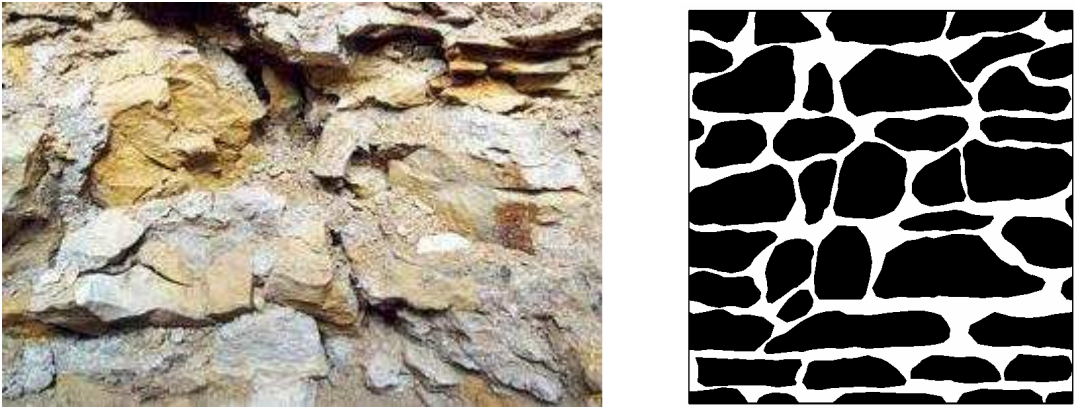

(c)

Figure 2: Typical masonry morphologies and corresponding unit cells: a) regular periodic masonry, b) non-periodic texture, c) irregular quarry filling.

Once an appropriate geometrical model, the PUC, of a given material system is specified, the subsequent homogenization analysis can be executed for each system independently. Owing to space limitations, in the subsequent sections we focus on the most complicated system, quarry masonry, shown in Fig. 2(c).

# 3 Local material parameters: mixed experimental and numerical modelling and model calibration

Unlike material parameters of individual constituents, which can be derived from conventional laboratory experiments, the estimates of model parameters along the common interface represent a rather delicate task. The problem becomes particularly important realizing the quasi-brittle character of both the mortar and stone phase, manifested by strain softening which emerges once the tensile strength has been exceeded. This leads (in dependence on the fracture toughness of stone and mortar) to the localization of inelastic strains mainly into the mortar joints between stone blocks.

A reasonably simple approach allowing for the prediction of material parameters of the interface transition zone (ITZ) relies on an appropriate numerical-experimental analysis. There is a variety of techniques of how to optimize the input data. One approach is very simple and starts from a set of input parameters based on the "trial and error" procedure. The calculated loading path is compared with that obtained experimentally. The least square method applied to minimize the difference between the calculated and measured loading force (the test is controlled by displacement) then yields the optimized model data. Another way, adopted in the present contribution, stands to benefit from sets of randomly generated input data using Monte Carlo or LHS Sampling methods. The best choice of the optimized input data again takes advantage of the least square method.

**3.1 Description of material test**

When examining heterogeneous materials it is rather difficult to reach the correspondence of the results obtained experimentally and using computational simulations. As for quarry masonry the problem consists in both the common texture of random character and somewhat vague material properties of the interfacial transition zone between the mortar joints and stone blocks. Lower values of cohesion and tensile strength are mainly affected by two phenomena: First, suction in the pores of dry blocks changes the water content in mortar causing incomplete hydration of the bonding agent. The second detrimental source of impaired contact properties are air bubbles in the pores of dry stone blocks. This is the reason why the strength of the masonry made up of blocks saturated with water before blocklaying (i.e. with no bubbles in the pores of blocks) distinctively increases when compared with the masonry made up of dry blocks.

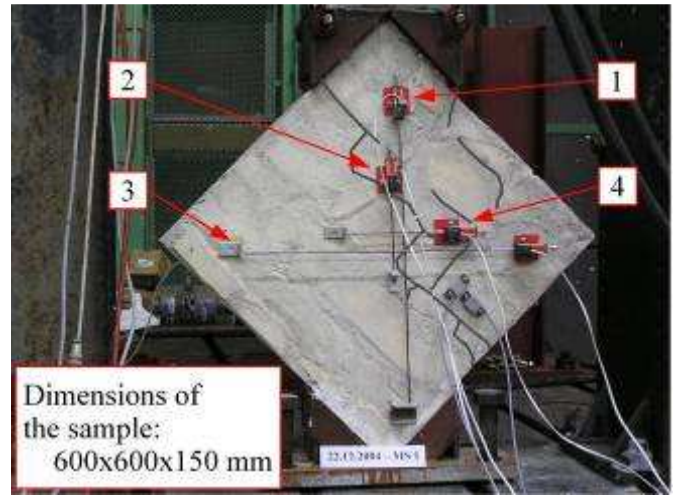

Figure 3: Crushing test of a quarry masonry sample

All the reported computational simulations are performed similarly to our proceeding works with the aid of the ATENA commercial software utilizing a plastic-fracturing NonLinearCementitious model exploiting the mesh-adjusted softening modulus in the smeared-crack approach to avoid the mesh dependent results (Červenka et al., 2002). The code allows us to account for the reduced cohesion and tensile strength of the ITZ by means of contact elements. The material model assumes the Mohr-Coulomb criterion with a corresponding yield surface cut-off by a tensile and compressive cap. The random texture of masonry generally tends to be described by means of the SEPUC (Povirk, 1995; Zeman and Šejnoha, 2007). However, in this section the actual image of the sample used for the experimental examination has been subjected to the finite element discretization, see Fig. 3.

Loading in compression was selected because in this particular case the satisfactory correspondence of the computationally obtained results with experimental outputs is rather difficult to achieve. The load vs strain diagrams obtained by the loading test are displayed in Fig. 4 and serve as bases for the calibration of the computational model. The positions of strain gauges 1-4 are evident from Fig. 3.

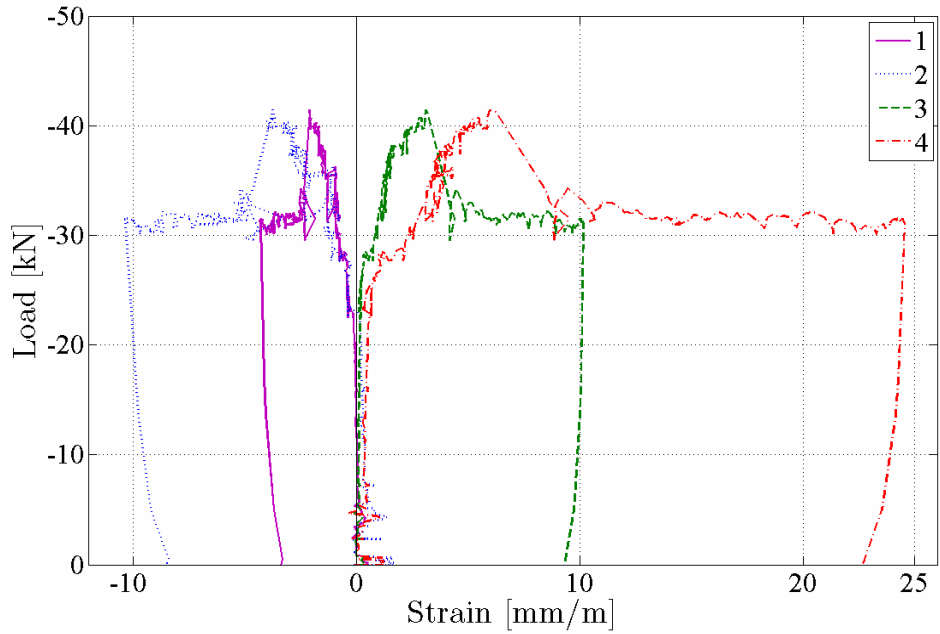

Figure 4: Loading paths: Load vs strain ($\varepsilon_1$ to $\varepsilon_4$) curves

A detailed finite element discretization is displayed in Fig. 5. The proposed model with contact elements along the contours of individual stone blocks is depicted in Fig. 5(a). Fig. 5(b) shows simplified mesostructural modelling in which the blocks are enlarged up to the middle surface of the mortar joint and the interconnection between the two adjacent blocks is performed by contact elements of zero thickness.

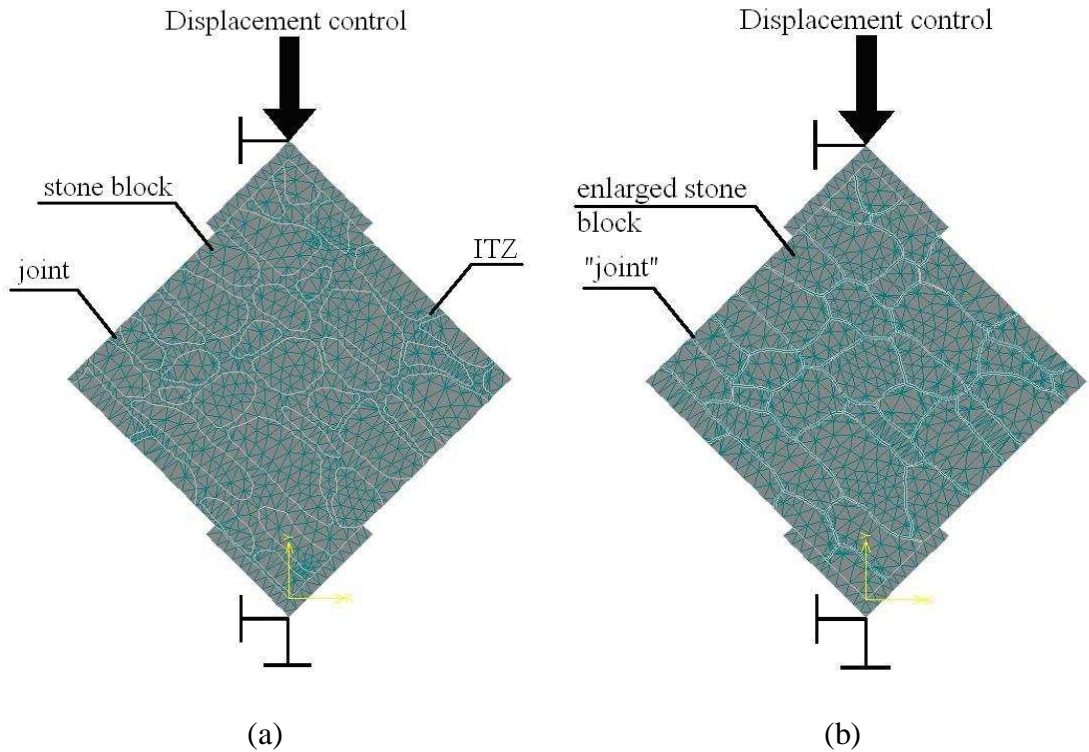

(a)                                            (b)

Figure 5: Finite element mesh of a quarry masonry sample: a) enhanced model, b) simplified model

**3.2 Model verification and results of solution**

Two variants of the solution are applied to solve this problem. The plots presented in Fig. 6 pertain to the enhanced model with contact elements situated along the boundary between the stone blocks and mortar (see Fig. 5(a)). Similar results, based on the trial and error strategy, were obtained by applying the simplified approach and substituting the real mortar texture by a set of expanded stone blocks interconnected with contact elements (see Fig. 5(b)). The material parameters of individual components of the model are summarized in Table 1. Note that the quasi-brittle characteristics of stone blocks and mortar were determined from a series of experimental tests executed at the Klokner institute, CTU in Prague (Novák et al. 2006), and kept constant during the identification procedure. The parameters of the interface, on the other hand, are calibrated using an experimental-numerical approach.

| Stone | | | | |
|---|---|---|---|---|
| $E$ [GPa] | $\nu$ [-] | $f_c$ [MPa] | $f_t$ [MPa] | $G_F$ [N/m] |
| 20.21 | 0.16 | 71.02 | 8.34 | 85.50 |

| Mortar | | | | |
|---|---|---|---|---|
| $E$ [GPa] | $\nu$ [-] | $f_c$ [MPa] | $f_t$ [MPa] | $G_F$ [N/m] |
| 5.30 | 0.18 | 6.10 | 1.31 | 6.70 |

| ITZ | | |
|---|---|---|
| $c$ [MPa] | $\varphi$ [-] | $f_t$ [MPa] |
| 0.13 | 0.30 | 0.10 |

Table 1: Material parameters of individual components

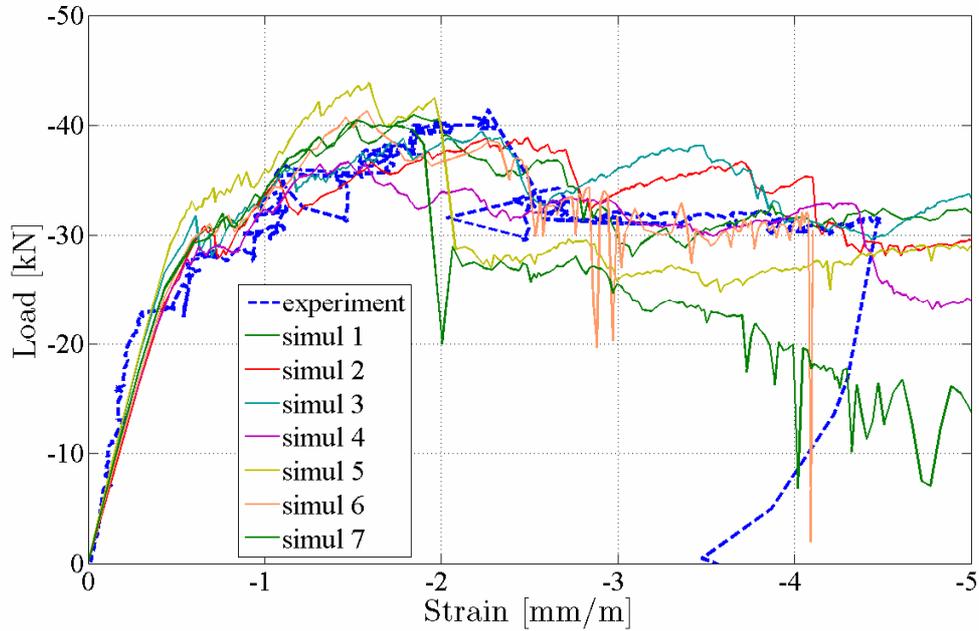

Figure 6: Calibration of the enhanced model (random population)

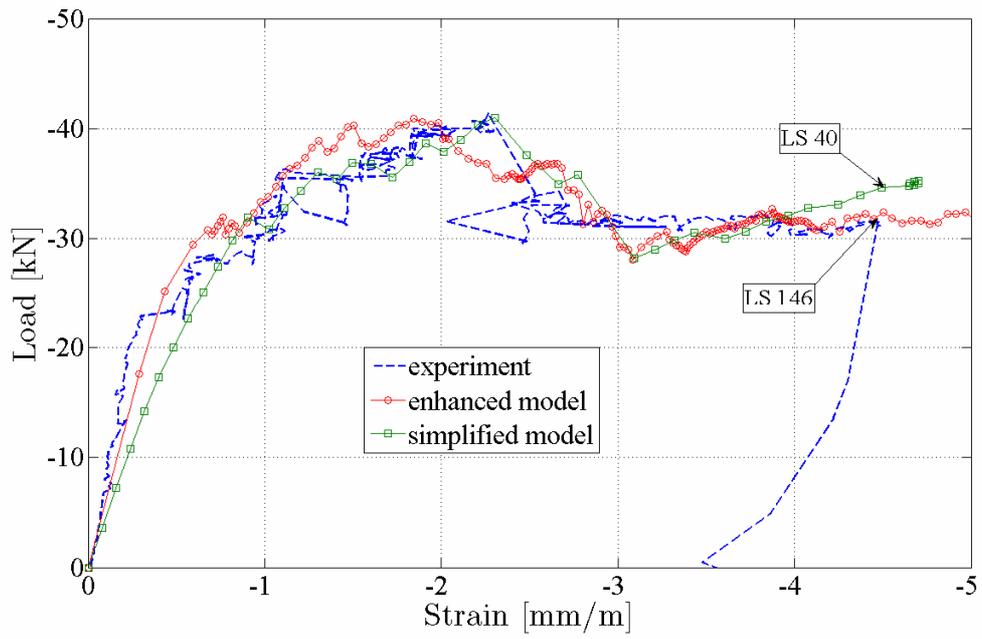

Figure 7: Verification of both model variants

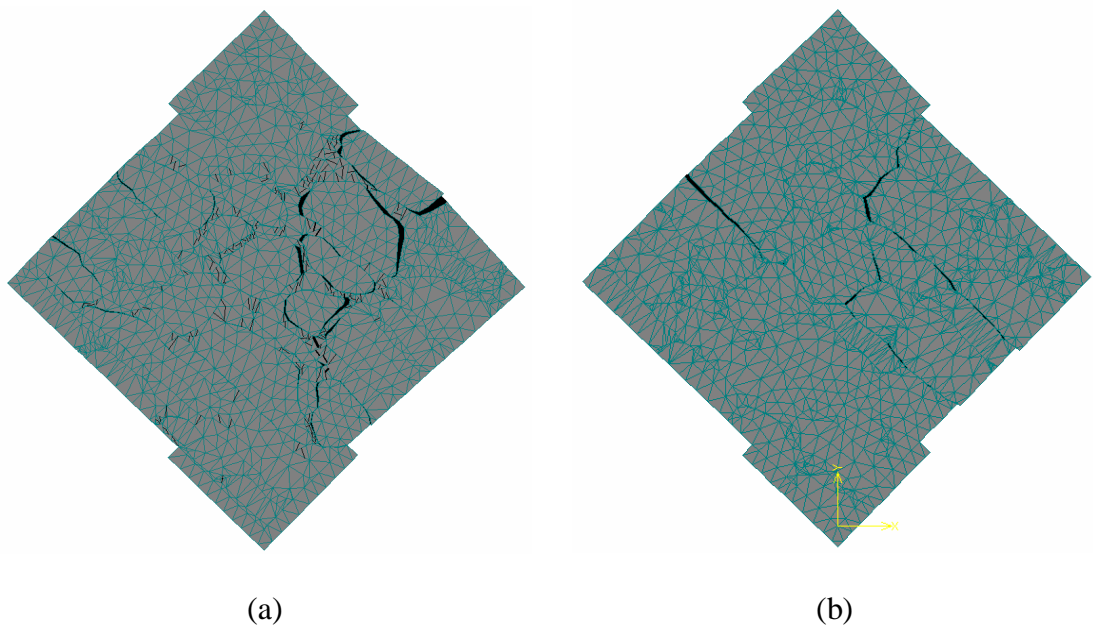

(a)                                                 (b)

Figure 8: Distribution of cracks pertinent to the maximum of experimentally prescribed strain: a) enhanced model (loading step 146), b) simplified model (loading step 40), magnified twice

To select the optimal solution, the least square method is applied to the objective function in this simple form

$$\int_0^{\varepsilon_{max}} \{[F(\varepsilon, p_i)] - \overline{F}(\varepsilon)\}^2 \, d\varepsilon = \min. \tag{1}$$

In Eq. (1), $p_i$, $i = 1, 2, 3,$ represent three material parameters of the contact, i.e. the cohesion $c$, the tensile strength $f_t$ and the angle of internal friction $\varphi$. These parameters fundamentally influence the dependence of the applied force on the prescribed displacement of the upper support, see Fig. 3. Considering the arrangement of the experimental test, the strain $\varepsilon_1$ (Fig. 3 and 4) has been inserted into Eq. (1) instead of the absolute displacement of the upper support. To reach a better agreement of the predicted loading path with that obtained experimentally, especially in the elastic phase, it is expedient to adjust by the back analysis also some of the material parameters pertaining to basic materials, e.g. the Young modulus of stone blocks.

The optimized load-strain diagrams obtained by both model variants are compared with experimental results, in particular with the strain measured in the loading direction in Fig. 7. The distribution of cracks, which corresponds to the maximum experimentally prescribed strain, is shown in Fig. 8 again for both variants of the computational model. The experimentally observed crack pattern is depicted in Fig. 3 for further comparison.

# 4 Prediction of effective properties in quarry masonry

Estimates of the ultimate load bearing capacity of historical structures often require a complex nonlinear full scale analysis. Clearly, introducing all geometrical details of the meso-structure within a macroscopic computational model would be prohibitively expensive. The crucial step thus appears in the derivation of the estimates of macroscopic or homogenized effective properties (Torquato, 2002; Milani, 2004).

In particular, having derived the relevant local material parameters this task is accomplished with the help of the first-order homogenization technique based on periodic fields. The theoretical formulation is briefly reviewed in Section 4.1. With reference to the application of periodic fields, no objections represented by the PUC are expected when estimating the elastic effective properties, see e.g. (Michel et al., 1999). However, when it comes to material parameters describing failure one may argue that the concept of homogenization based on periodicity assumptions and the existence of uniform fields is objectionable, especially if dealing with quasi-brittle materials prone to localized rather than distributed damage. In the present approach, however, when the analysis on two relevant scales is totally uncoupled, the homogenized properties are introduced directly into the macroscopic constitutive law. As no back reference to the actual heterogeneous meso-structure is made, the evolution of a highly localized failure zone due to strain softening is correctly captured by the macroscopic model. As an example of this approach, in Section 4.2 we present numerical predictions of macroscopic fracture energy. Section 4.3 finally provides validation of the applicability of the homogenization technique by comparing the results derived from the periodic unit cell analysis and those found from a numerical simulation based on the concept of the macroscopic wedge splitting test often used in experimental determination of macroscopic fracture energy (RILEM, 1985; Brühwiler and Wittmann, 1990; Bažant and Kazemi MT, 1991; Šejnoha et al., 2006).

## 4.1 First-order homogenization - theoretical formulation and boundary conditions

Consider a heterogeneous periodic cell $Y$ subjected to a uniform macroscopic strain $E$. In view of the periodicity of the cell, the strain and displacement fields in the PUC admit the following decomposition

$$u(x) = E \cdot x + u^*(x), \qquad \varepsilon(x) = E + \varepsilon^*(u^*(x)). \qquad (2)$$

The first term in Eq. (2a) corresponds to a displacement field in an effective homogeneous medium which has the same overall response as the composite aggregate; see e.g. (Michel et al.; 1999) and references therein. The fluctuating $Y$-periodic displacement $u^*$ and the corresponding strain $\varepsilon^*$ enter Eq. (2) as a consequence of the presence of heterogeneities. Note that the periodicity of $u^*$ further implies that the average of $\varepsilon^*$ in the unit cell vanishes. The local stress fields $\sigma$ in the PUC are constrained by equilibrium conditions

$$\partial^T \sigma(x) = 0, \qquad (3)$$

together with appropriate constitutive laws. Note that the symbol $\partial^T$ denotes the "equilibrium" operator matrix (Bittnar and Šejnoha, 1996). Combining Eqs. (3) and (2) and the stress-to-strain map allows us to determine the distribution of the fluctuating displacement $u^*$ within the cell as a function of $E$ and subsequently to evaluate the macroscopic average stress $\Sigma$ in the PUC. This procedure yields the homogenized constitutive relation in the form

$$\Sigma = \frac{1}{|Y|} \int_Y \sigma(u^*(E)) dY. \qquad (4)$$

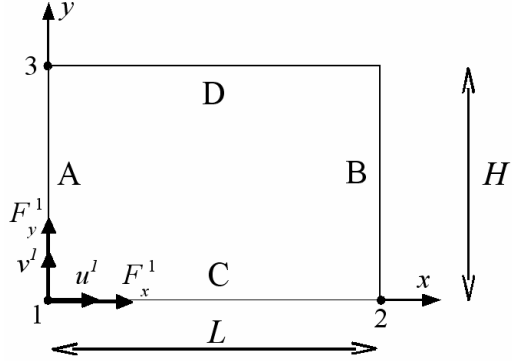

Figure 9: Scheme of a PUC and controlling points

In the present work, the Finite Element-based homogenization is based on the concept of controlling points introduced by Teplý and Dvořák (1998) see also (Kouznetsova et al., 2001; Massart et al., 2007). In this context, the macroscopic strain load $E$ is imposed on the cell by prescribing the values of displacements $u$ and $v$ in points 1, 2, and 3 as indicated in Fig. 9. For this particular choice, the following relation between the controlling displacements and the macroscopic strain components holds:

$$\begin{Bmatrix} u^1 \\ v^1 \\ u^2 \\ v^2 \\ u^3 \\ v^3 \end{Bmatrix} = \begin{bmatrix} 0 & 0 & -H \\ 0 & 0 & 0 \\ L & 0 & -H \\ 0 & 0 & 0 \\ 0 & 0 & 0 \\ 0 & H & 0 \end{bmatrix} \begin{Bmatrix} E_{xx} \\ E_{yy} \\ 2E_{xy} \end{Bmatrix}, \qquad (5)$$

where $H$ and $L$ are the dimensions of the rectangular PUC shown in Fig. 9. The periodic character of the fluctuating part of the displacement $u^*$, recall Eq. (2), is introduced using linear constraints between the homologous edges of the unit cell:

$$\begin{aligned} \boldsymbol{u}^B &= \boldsymbol{u}^A + \boldsymbol{u}^2 - \boldsymbol{u}^1, \\ \boldsymbol{u}^D &= \boldsymbol{u}^C + \boldsymbol{u}^3 - \boldsymbol{u}^1, \end{aligned} \qquad (6)$$

which can easily be introduced in the majority of commercial codes.

Finally, using the equilibrium conditions on the corresponding edges of the PUC, the values of the macroscopic stresses can be directly extracted from the reaction forces acting on the selected controlling points (see (Teplý and Dvořák, 1998; Kouznetsova et al., 2001; Massart et al., 2007) for further details)

$$\begin{Bmatrix} \Sigma_{xx} \\ \Sigma_{yy} \\ \Sigma_{xy} \end{Bmatrix} = \frac{1}{tHL} \begin{bmatrix} 0 & L & 0 \\ 0 & 0 & H \\ -H & -H & 0 \end{bmatrix} \begin{Bmatrix} F_x^1 \\ F_x^2 \\ F_y^3 \end{Bmatrix}, \qquad (7)$$

where $t$ is the thickness of the PUC.

### 4.2 Prediction of macroscopic fracture energy from homogenization

A methodology for evaluating the size independent fracture energy $G_F$ from homogenization was discussed in detail in [Zeman and Šejnoha, 2007; Šejnoha et al., 2006]. It was shown that, in the case of a straight crack perpendicular to the principal strain, $E_{xx}$, this quantity can be expressed as the area under the macroscopic stress-strain curve displayed in Fig. 10(a) and multiplied by the length of the periodic unit cell $L$ as

$$G_F^x = \int_0^{W^c} \Sigma_{xx} dW^c = \int_0^{f_t} (E_{xx} - E_{xx}^{el}) L d\Sigma_{xx} = L \int_0^{E_{\max}} \Sigma_{xx} dE_{xx}, \qquad (8)$$

where $W^c$ is the assumed macroscopic crack opening displacement.

Generalization of Eq. (8) suitable to a more complicated crack pattern of Fig. 11 (typical of quarry masonry), and yet consistent with the RILEM work-of-fracture relation (11) introduced in (RILEM, 1985), reads

$$G_F^x = \frac{\int_0^{u_{max}} F(u)\,du}{A_{crack}} = \frac{BH \int_0^{E_{xx,max}} \Sigma_{xx}\,d(E_{xx}L)}{aB} = \frac{LH}{a}\int_0^{E_{xx,max}} \Sigma_{xx}\,dE_{xx}, \qquad (9)$$

where $a$ represents the total length of traction free surfaces.

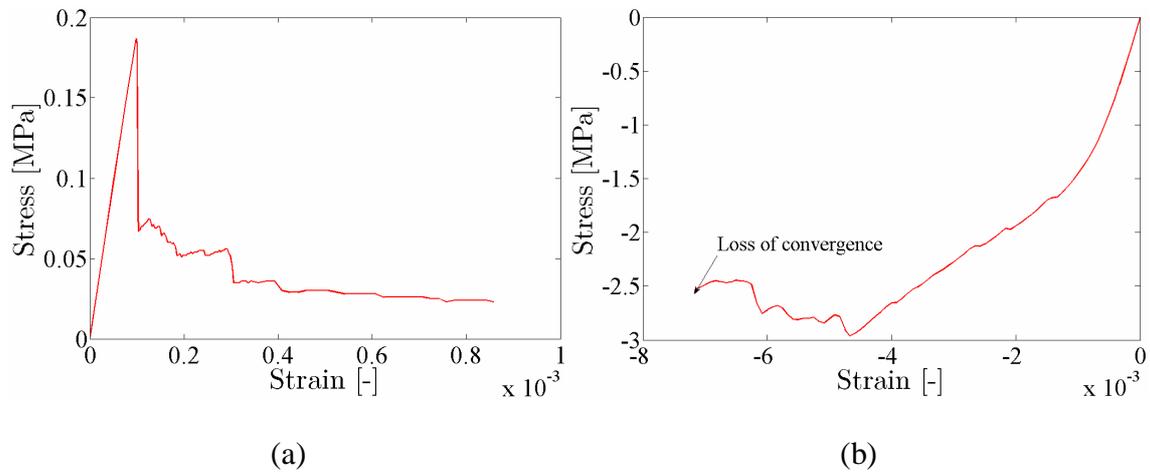

(a)            (b)

Figure 10: PUC analysis: Macroscopic stress-strain curves: a) tensile loading, b) compressive loading

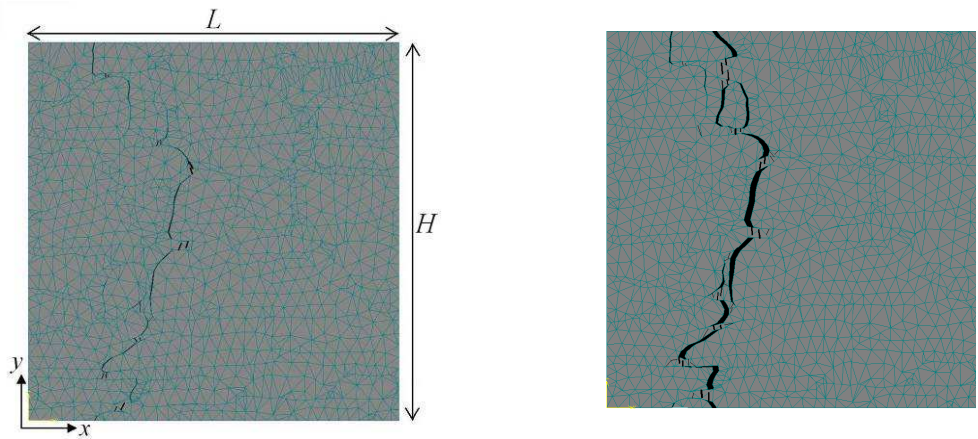

(a) (b)

Figure 11: PUC analysis: Distribution of cracks for irregular arrangement of blocks: a) onset of cracking, b) ultimate failure

A similar approach can also be employed for the $yy$-direction. Providing the two values of fracture energies $G_F^x$ and $G_F^y$ vary due to their inherited orthotropic behaviour one may accept the single value fracture energy to be introduced in the macroscopic constitutive law, generally assumed isotropic, in the form

$$G_F = \sqrt{G_F^x G_F^y}. \tag{10}$$

Note that the elastic limit point on the stress strain curve in Fig. 10(a) represents the homogenized (macroscopic) tensile strength. The results of the compression test plotted in Fig. 10(b) then serve to extract the corresponding macroscopic compressive strength.

**4.3 Effective fracture energy from macroscopic simulations - wedge splitting test**

The specific fracture energy $G_f(W, W/a)$ determined as the total work of fracture divided by the projected fracture area

$$G_f(W, W/a) = \frac{1}{(W-a)B} \int_0^{u_{max}} F(u)\,du, \tag{11}$$

may experience, owing to the variation of the fracture process zone, a certain size dependence (RILEM, 1985; Brühwiler and Wittmann, 1990). Recall that parameters $W$, $a$ and $B$ in Eq. (11) represent the specimen depth, the initial crack length (wedge depth) and the specimen thickness, respectively, see also Fig. 12(a). The specific fracture energy $G_f$ can be expressed as the mean value of the local fracture energy $g_f(x)$ as

$$G_f(a) = \frac{1}{W-a} \int_0^{W-a} g_f(x)\mathrm{d}x \leq g_f(a). \tag{12}$$

Assuming a bilinear form of $g_f$ depicted in Fig. 12(c), the relationship between $G_f$ and the size independent fracture energy $G_F$ is given by, see (Duan et al., 2003)

$$G_f(a) = G_F \frac{W-a}{2a_l}, \qquad a \geq W - a_l \tag{13}$$

$$G_f(a) = G_F \left[1 - \frac{a_l}{2(W-a)}\right], \qquad a \leq W - a_l, \tag{14}$$

where $a_l$ stands for the transition ligament size, see Fig. 12(c). The least square method is usually called for to derive the two unknown parameters $G_F$ and $a_l$ from a series of tests with a different notch to depth ratio as long as $(W-a) > a_l$.

Moreover, (Karihaloo et al., 2003) promoted the possibility of deriving the two unknown parameters from a single size specimen with only two notches to depth ratios providing they are well separated. This particular option was also examined in the present study. Nevertheless, four specimens with variable notch to depth ratios, Fig. 12(b), were analyzed first to confirm the variation trend of the size dependent fracture energy $G_f$, Fig. 12(c), together with the applicability of Eqs. (13) and (14).

The finite element discretization of two specific samples with the smallest and the largest notch to depth ratio appears in Fig. 13. The complicated crack patterns for the two specimens are displayed in Fig. 14.

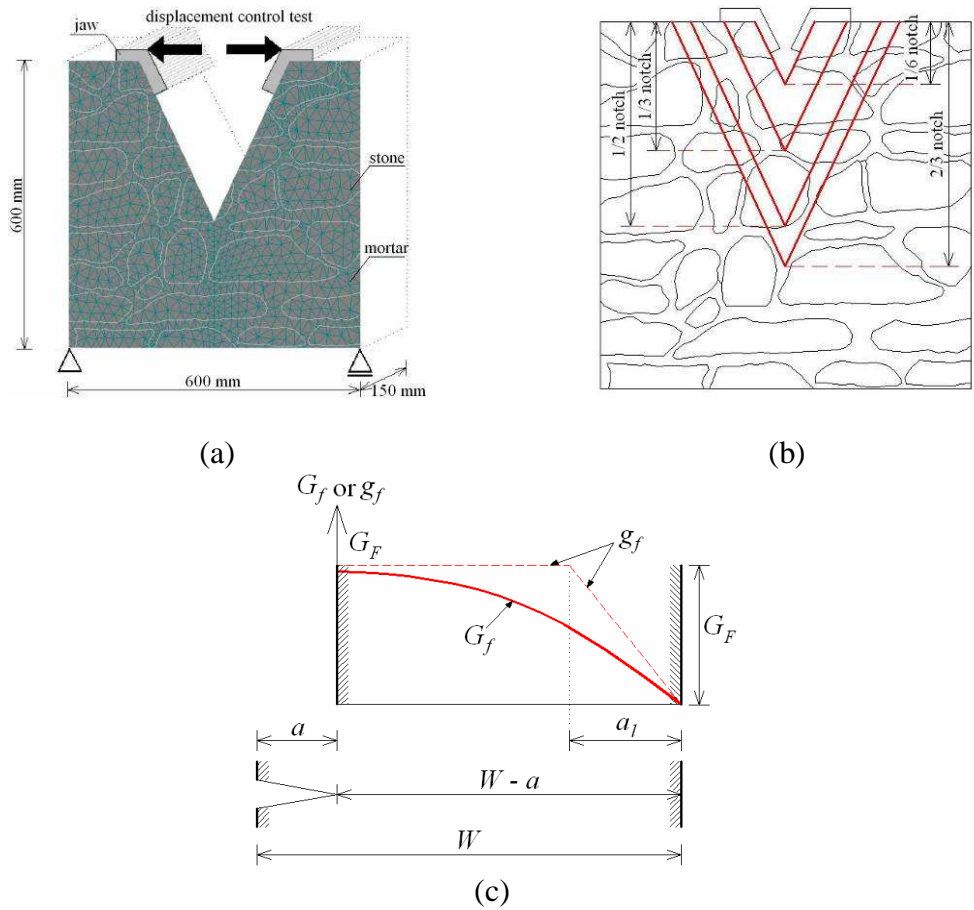

Figure 12: Wedge splitting test: a) experimental setup, b) view of selected notches, c) graphical representation of fracture energies

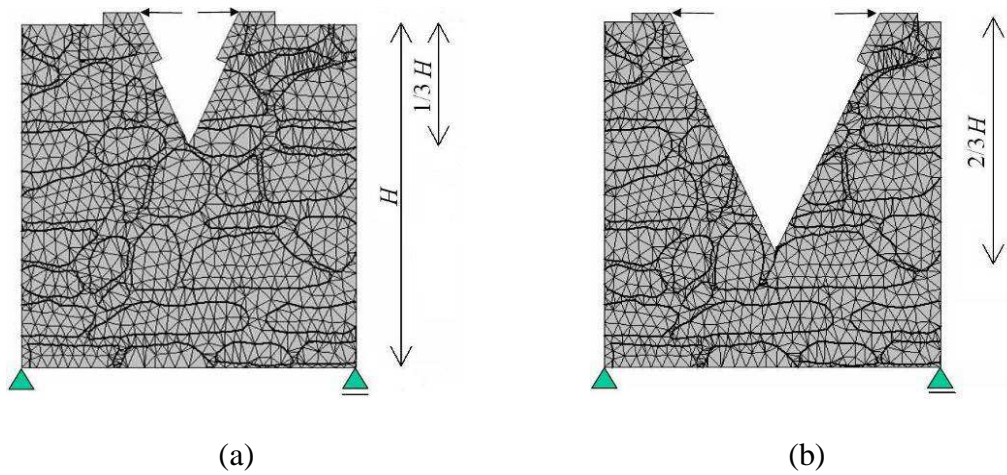

Figure 13: Wedge splitting test: Geometry and finite element mesh: a) shallow notch, b) deep notch

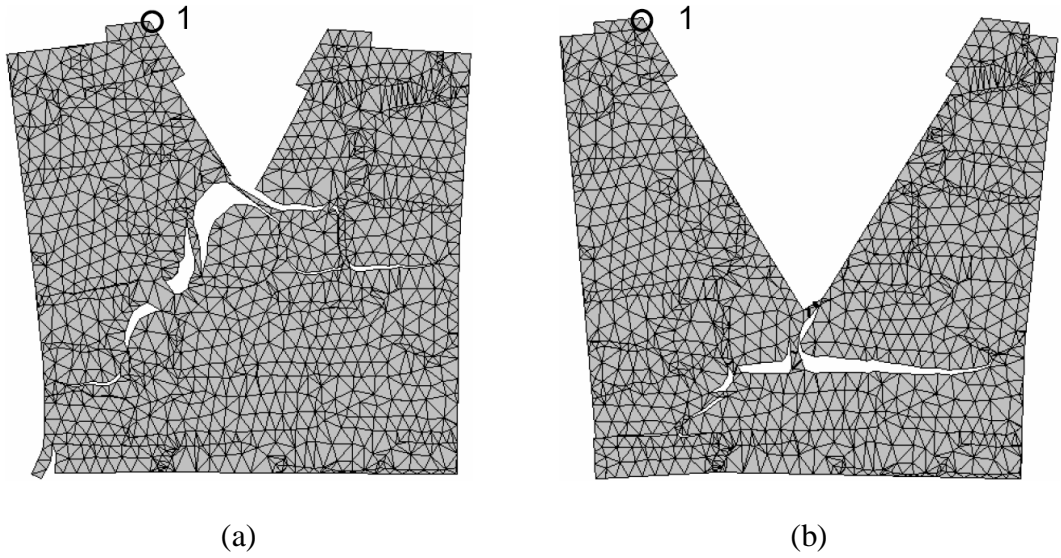

(a) (b)

Figure 14: Wedge splitting test: Distribution of cracks: a) shallow notch, b) deep notch

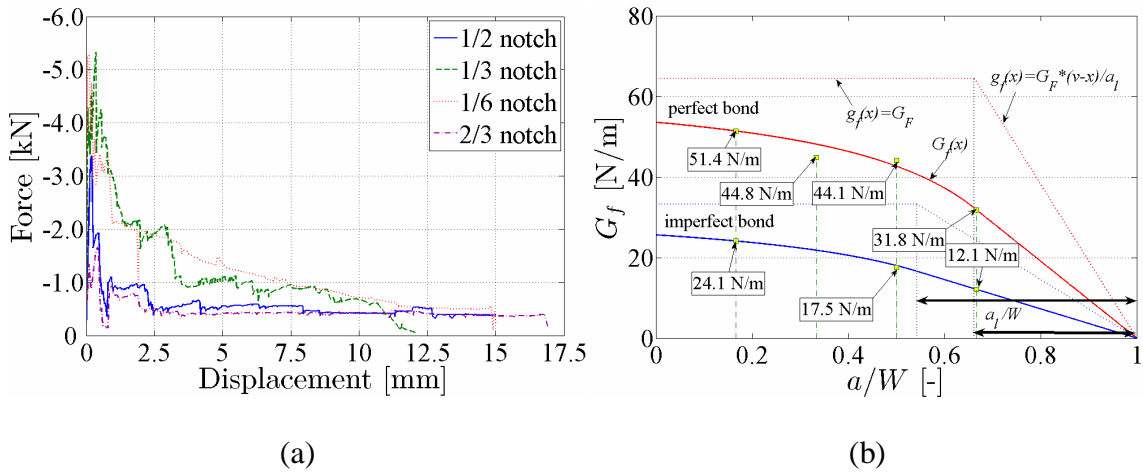

(a) (b)

Figure 15: Wedge splitting test: a) macroscopic response, b) estimated fracture energy

It is worth noting a rather unrealistic long tail in the force-displacement diagram (see Fig. 15) for some of the specimens. The final plot of fracture energies $G_f$ presented in Fig. 15(b) also demonstrates the quality of the results noting that numerically obtained values of individual fracture energies $G_f$ follow the expected trend reasonably well. Finally, the two values for the smallest and the largest notch to depth ratios were introduced into Eqs. (13) and (14) to estimate the size independent fracture energy $G_F$ and the transition ligament size just to check the imposed constrain condition, $(W-a) > a_l$.

To account for the effect of the ITZ, the above analysis was carried out again, but this time the computational model was enhanced by introducing contact elements along the stone block-mortar interfaces. The results in Fig. 15(b) clearly show the expected drop of the size independent fracture energy $G_F$.

The values of fracture energies derived from both approaches considering perfect as well as imperfect bonds between the stone block and mortar phases are stored in Table 2. A good agreement between both methods is self-evident thus supporting the applicability of the homogenization technique even in applications involving strain-softening materials.

| Wedge splitting test | | Homogenization | |
|---|---|---|---|
| perfect bond | imperfect bond | perfect bond | imperfect bond |
| 64 | 33 | 67 | 31 |

Table 2: Size independent fracture energies $G_F$ [N/m]

# 5 Discussion and conclusions

In order to realistically model masonry structures with complex geometries, the non-linear response of individual components must inevitably be taken into account as described in Sections 3 and 4.

In Section 3 the efficiency of a detailed computational model for quarry masonry has been compared with that of a simplified model, in which the contacts between the adjacent stone blocks are conveyed by contact elements of zero thickness. These elements are placed onto the middle line (in 2D modelling and/or the middle surface in 3D modelling) of the mortar bed. The respective stone blocks are expanded up to this boundary. The improved model combines the finite elements when discretizing the mortar joints with contact elements to cover the impaired material properties of the ITZ. Fig. 7 suggests that both models are viable and applicable in engineering practice. Point out that the augmented model shows certain merits: (i) a better agreement of the computationally predicted response with that obtained experimentally, both in the elastic region as well as in the state with fully developed cracks (plateau of the load – strain diagram preceding the collapse of the analyzed sample); (ii) the back analysis is restricted to a small set of relatively close and only slightly scattered curves; (iii) the image of cracks predicted by computational simulations seems to be a better approximation to the real pattern of cracks, compare Fig. 3 and Fig. 8.

Someone may object that, from the practical point of view, the differences between the presented results are insignificant. This opinion may perhaps be accepted in the case of mechanical loading. On the other hand, no simplification of this kind can be made when analysing transport processes in masonry considering heat and moisture fluxes across the ITZ. In this very important case an imperfect hydraulic contact on the interface manifests itself by different pore size distributions of the adjacent porous materials which results in a jump in capillary pressures (Černý and Rovnaníková, 2002). It is plain enough to expect that the jump in capillary pressures yields a corresponding jump in the temperature field. This easily follows from the application of weak formulation to the heat balance equation including convective terms. This problem will be discussed in detail in a forthcoming journal paper.

Clearly, both models are applicable to the non-linear homogenization of effective (macroscopic) material properties, which are necessary for the integrity assessment of masonry structures. In Section 4, two specific approaches to the derivation of macroscopic (effective) fracture energy needed in full scale macroscopic simulations were examined. The first approach exploits the well known elements of the first order homogenization in conjunction with the statistically equivalent periodic unit cell, while the second approach draws on the numerical representation of standard laboratory tests proposed for the determination of size independent fracture energy for quasi-brittle materials including concrete and masonry structures. A comparison of the results suggests a good agreement between individual approaches and therefore their applicability for the present problem. Owing to its relative simplicity over the more tedious wedge splitting test, the former approach appears to be the more efficient one particularly in the case of virtual (numerical) experiments.

The macroscopic material data obtained in the way described in Sections 3-4, was used in the ATHENA 3D code to perform a detailed three dimensional analysis of Charles Bridge in Prague (Zeman et al., 2006).

**Acknowledgement**

This outcome has been achieved with financial support of the Ministry of Education, Youth and Sports project No. 1M680470001, within activities of the CIDEAS research centre. In this undertaking, theoretical results gained in the project GAČR 103/04/1321 were partially exploited.

**List of figures**

Figure 1: a) A view of a typical bridge arch showing a regular arrangement of stones in masonry, b) a crack running both across stone and along a head joint

Figure 2: Typical masonry morphologies and corresponding unit cells: a) regular periodic masonry, b) non-periodic texture, (c) irregular quarry filling.

Figure 3: Crushing test of a quarry masonry sample

Figure 4: Loading paths: Load vs strain ($\varepsilon_1$ to $\varepsilon_4$) curves

Figure 5: Finite element mesh of a quarry masonry sample: a) enhanced model, b) simplified model

Figure 6: Calibration of the enhanced model (random population)

Figure 7: Verification of both model variants

Figure 8: Distribution of cracks pertinent to the maximum of experimentally prescribed strain: a) enhanced model (loading step 146), b) simplified model (loading step 40), magnified twice

Figure 9: Scheme of a PUC and controlling points

Figure 10: PUC analysis: Macroscopic stress-strain curves: a) tensile loading, b) compressive loading

**List of tables**

**Keywords**

meso-scale, macro-scale, statistical descriptor, fracture energy, masonry.